\documentclass[%
 reprint,
 amsmath,amssymb,
 aps,
prb,
]{revtex4-1}

\usepackage{graphicx}
\usepackage{epstopdf}
\usepackage{dcolumn}
\usepackage{bm}
\usepackage{gensymb}
\usepackage{upgreek}

\begin{document}

\preprint{APS/123-QED}

\title{CoAs: The line of 3\textit{d} demarcation}

\author{Daniel J. Campbell$^1$}
\author{Limin Wang$^1$}
\author{Chris Eckberg$^1$}
\author{Dave Graf$^2$}
\author{Halyna Hodovanets$^1$}
\author{Johnpierre Paglione$^{1,3}$}
\affiliation{%
 $^1$Center for Nanophysics and Advanced Materials, Department of Physics, University of Maryland, College Park, Maryland 20742, USA\\
 $^2$National High Magnetic Field Laboratory, 1800 East Paul Dirac Drive, Tallahassee, Florida 32310, USA\\
 $^3$Canadian Institute for Advanced Research, Toronto, Ontario M5G 1Z8, Canada
}%

\date{\today}

\begin{abstract}

Transition metal-pnictide compounds have received attention for their tendency to combine magnetism and unconventional superconductivity. Binary CoAs lies on the border of paramagnetism and the more complex behavior seen in isostructural CrAs, MnP, FeAs, and FeP. Here we report the properties of CoAs single crystals grown with two distinct techniques along with density functional theory calculations of its electronic structure and magnetic ground state. While all indications are that CoAs is paramagnetic, both experiment and theory suggest proximity to a ferromagnetic instability. Quantum oscillations are seen in torque measurements up to 31.5~T, and support the calculated paramagnetic Fermiology.

\end{abstract}

\maketitle

\section{\label{sec:Intro}Introduction}

Materials composed of transitions metals and pnictogens have been the subject of much recent study due to their interesting physical properties. The relationship between magnetic ordering, structural changes, and high-temperature superconductivity has been a focus of research in the iron-based superconductors\cite{PaglioneFeSCs, DaiFeSCs}. Binary compounds of Nb, Ta, As, and P have been identified as topological semimetals\cite{LvTaAs1, LvTaAs2, YangTaAs, Di-FeiNbP, ShekharNbP, XuSYTaP, XuNTaP, XuSYNbAs}. Other 3\textit{d} metal-pnictide binary materials crystallizing in the orthorhombic, MnP-type \textit{Pnma} structure show complex magnetic states, and in some cases superconductivity. From left to right on the periodic table, CrAs and MnP are helimagnets at low temperature, but superconduct as the magnetic transition is driven away with applied pressure\cite{KotegawaCrAs, WuCrAs, ChengMnP}. MnP is particularly complex, with multiple ferromagnetic (FM) and antiferromagnetic (AFM) orderings as a function of temperature and pressure\cite{ChengCrAsMnP}, while replacing the pnictogen to form MnAs yields a room-temperature ferromagnet\cite{SaparovTAs}. FeP shows a similar ordering to CrAs and MnP\cite{KallelHelimagnetism}, though it has not yet been found to superconduct. Binary FeAs has an unusual non-collinear spin-density wave transition at low temperatures\cite{RodriguezFeAsSDW} and offers a compositional link to the high T$_c$ iron superconductors. However, adding one more electron breaks the trend, as CoAs has shown no indication of magnetic ordering\cite{SelteCoAs, SaparovTAs}. One more step to the right yields NiAs, which is hexagonal and similarly paramagnetic (PM)\cite{SaparovTAs}.

Despite standing on the borderline of magnetic ordering in 3\textit{d}-pnictide binaries, there have been no low temperature reports on CoAs single crystals. Additionally, while previous powder or polycrystalline measurements show low temperature paramagnetism, features such as a nonmonotonic temperature dependence of the magnetic susceptibility have been noted but not explained\cite{SelteCoAs, SaparovTAs}. CoAs merits investigation as a PM but potentially magnetically unstable comparison to the complicated ordering of the Cr, Mn, and Fe-based materials, which have been wellsprings of interesting physical phenomena.

In this paper we present two ways to make CoAs single crystals: chemical vapor transport (CVT) using iodine gas, as has been done before, and a new bismuth flux technique. Each possesses its own advantages. Quantum oscillations have been observed in torque measurements at high fields in flux-grown CoAs. We have also made density functional theory (DFT) calculations, which together with quantum oscillations data can give an experimentally verified Fermi surface picture. Combined with resistivity, Hall effect, heat capacity, and magnetization measurements we provide a comprehensive overview of the properties of binary CoAs. The compound is indeed paramagnetic as previous reports have indicated, even though calculations slightly favor a FM ground state. However, features of the temperature-dependent magnetic susceptibility and heat capacity indicate possible magnetic fluctuations, leading us to conclude that CoAs is a near ferromagnet.

\section{\label{sec:Crystal Growth}Crystal Growth}

CoAs single crystals were prepared in two ways. We first present a Bi flux technique, combining prereacted CoAs powder with bulk Bi (Puratronic, 99.999\%) in a 1:20 ratio in an alumina crucible and sealing the combination in a quartz ampule under partial pressure of argon gas. Bi has been used as a flux to prepare FeAs single crystals\cite{CampbellFeAs}, and has the advantage of not forming a compound with either Co or As, reducing the chance of forming alternate phases. Multiple temperature profiles were tried with little noticeable change in crystal quality. For the sample for which oscillations data were taken, the growth was heated at a rate of 50~\degree{}C per hour to 900~\degree{}C then cooled at 2~\degree{}C per hour to 500~\degree{}C, at which point the ampule was spun in a centrifuge to separate crystals from flux. The crystals are small and platelike, with typical dimensions of 0.4~$\times$~0.2~$\times$~0.1~mm$^3$. The axis perpendicular to the basal plane is always \textit{c}, the longest crystal axis, as determined by single crystal x-ray diffraction (XRD).

In general, crystals grow as rectangular plates in flux. However, growths with a maximum temperature of 1000~\degree{}C and spin temperature of 925~\degree{}C had a more hexagonal shape. Many compounds transition between the orthorhombic MnP structure and the hexagonal NiAs structure at high temperatures, including CoAs (T$_S \approx{} 975$~\degree{}C)\cite{SelteNiAsMnP}. Tremel \textit{et al.} theorized that the orthorhombic structure was more stable than the hexagonal one in the binaries only for $d^2$ to $d^6$ transition metals\cite{TremelNiAsMnP}. CoAs being orthorhombic at room temperature, despite its $d^7$ atom, bucks this trend and is another link to the Cr-Fe pnictides. Regardless of appearance, room temperature powder XRD of the hexagonally-shaped crystals shows them to be orthorhombic.

We have also grown crystals by chemical vapor transport with I$_2$, a technique frequently employed for CoAs and other 1:1 3\textit{d} metal arsenides\cite{RodriguezFeAsSDW, SelteCoAs, LymanCoAsPressure, SegawaFeAs, ZiqFeAsCVT, ZhuCrAsCVT, BinnewiesCVTBook, CampbellFeAs}. In this case the procedure is to combine prereacted CoAs powder with about 3~mg/mL of polycrystalline I$_2$ in an evacuated quartz ampule of length 15~cm and place the ampule in a horizontal tube furnace. A temperature gradient is maintained so that the end of the ampule containing the material is at 830~\degree{}C, while the empty end is at 600~\degree{}C. The powder initially at the hot end will react with the I$_2$ vapor and be transported in a gaseous state to the cold end. After approximately two weeks large crystals form at the cold end and the furnace is shut off. Any I$_2$ that condenses on the surface of the resulting crystals during cooldown is easily washed away with ethanol and does not manifest in later characterization measurements. CoAs forms readily through CVT and crystals are much larger than those grown out of flux, with dimensions exceeding 1~mm. A disadvantage is their irregular shape, making principal axes more difficult to identify. Compared to Bi flux crystals, they facilitate Hall effect, heat capacity, and magnetic measurements, at the cost of a less consistent orientation. Powder XRD measurements of CVT and flux crystals give the same lattice parameters: \textit{a}~=~5.28~\AA{}, \textit{b}~=~3.49~\AA{}, and \textit{c}~=~5.87~\AA{}, which also match previous results\cite{SelteCoAs, SaparovTAs}.

\section{\label{sec:Transport}Physical Properties}

\begin{figure}
    \centering
    \includegraphics[width=0.47\textwidth]{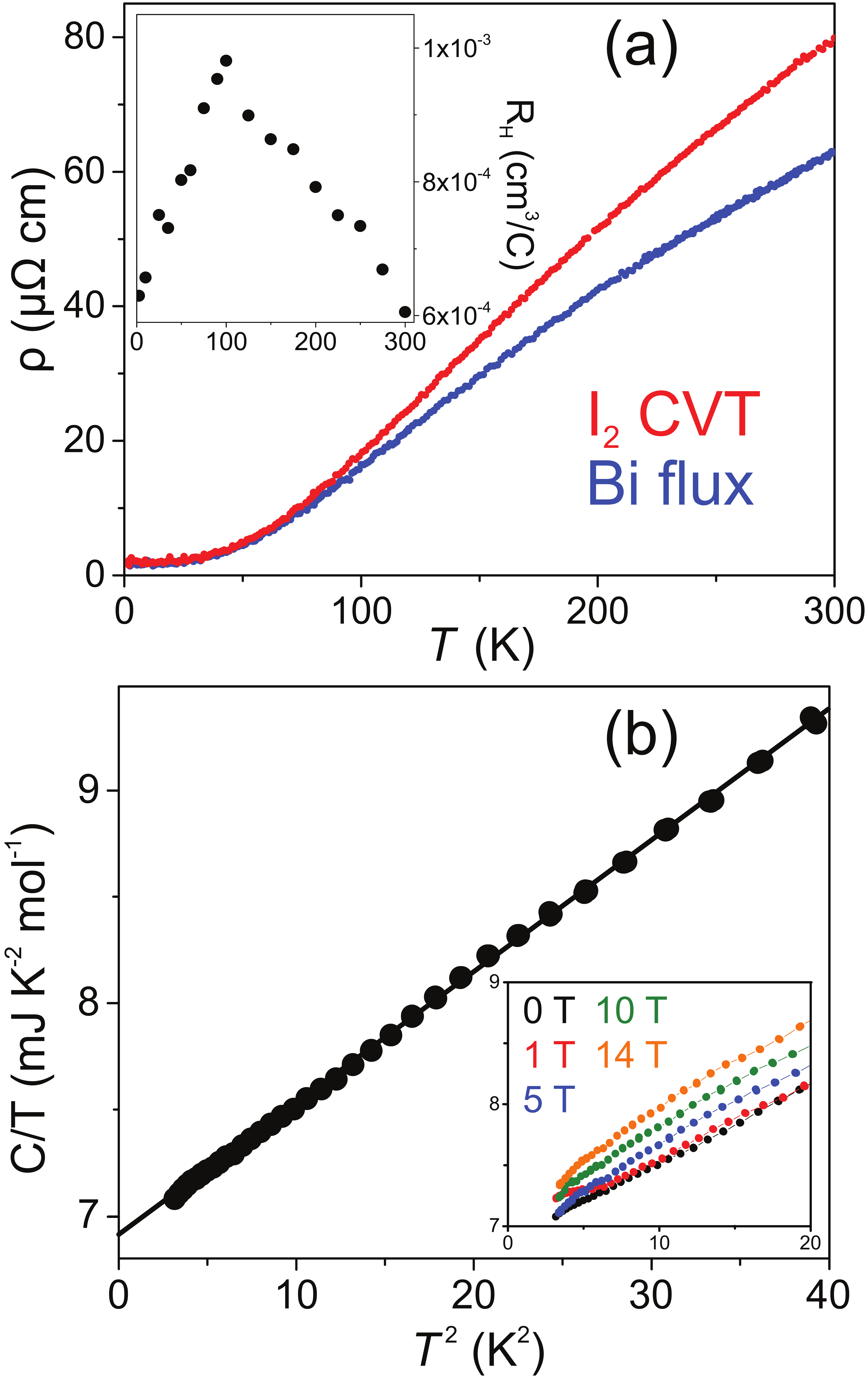}
    \caption{(a) The resistivity of CoAs single crystals grown by Bi flux and I$_2$ vapor transport. Inset: temperature dependence of the Hall coefficient. (b) Low temperature molar heat capacity. The line is a fit of the 0~T data to $C/T = \gamma{} + \beta{}T^2$ for 4.5~K~$<$~\textit{T}~$<$~7~K. Inset: a closeup of the same data, as well as that for various applied fields, at lowest temperature. Here, lines connect points and are not fits.}
    \label{fig:Figure1}
\end{figure}

The only previous reports of low temperature physical properties of CoAs have been on polycrystalline samples\cite{SaparovTAs}. The resistivity of single crystal CoAs [Fig.~1(a)] is 60-80~$\mu{}\Omega{}$~cm at room temperature and displays a featureless, slightly sublinear temperature dependence before saturating for T~$<$~30~K. The residual resistivity ratio (RRR), defined as $\rho$(300~K)/$\rho$(1.8~K), is up to 70 for flux crystals compared to about 40 for the best CVT samples. Typical resistivities at 1.8~K are roughly 1-2~$\mu{}\Omega{}$~cm, with slightly higher values for CVT samples. We interpret the higher RRR and lower resistivity of the smaller Bi flux crystals as an indication that they are of a higher quality than those grown with vapor transport. A higher RRR for Bi flux samples compared to vapor transport growths has also been seen for FeAs\cite{CampbellFeAs}, though the effect is not as dramatic for CoAs. As Fig.~1(a) indicates, residual resistivity values are still very close. Hall effect measurements were performed between $\pm{}$9~T on the wider vapor transport crystals, with the antisymmetric component of the Hall resistance used to calculate the Hall coefficient R$_H$ [Fig.~1(a), inset]. The antisymmetrized curves are linear with a positive slope over the entire temperature range, indicating hole-dominated conduction. Data sets for multiple samples show a peak near 100~K. Similar sharp extrema have been observed in R$_H$ measurements of Sb\cite{TanakaSb}, FeAs\cite{SegawaFeAs}, and CrB$_2$\cite{BauerCrB2}, and indicate the presence of multiple carriers with differing temperature dependences. The idea of multicarrier transport is also supported by theoretical calculations and quantum oscillations measurements which will be presented later on in this paper. In FeAs and CrB$_2$ the $R_H$ peaks occur at the onset of antiferromagnetism. However, no other measured properties of CoAs show features near the location of the R$_H$ maximum.

\begin{figure}
    \centering
    \includegraphics[width=.49\textwidth]{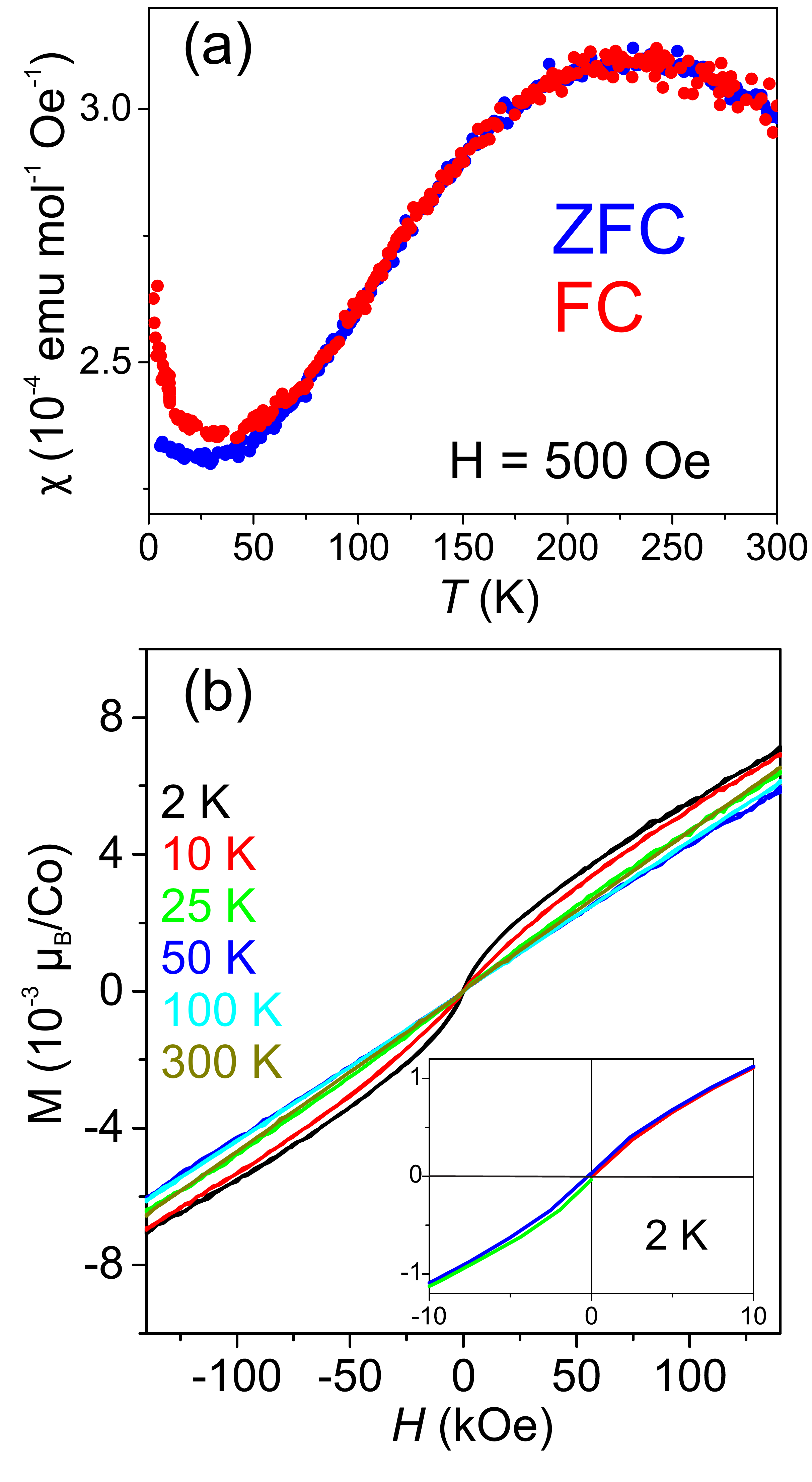}
    \caption{(a) Zero field cooled (ZFC) and field cooled (FC) magnetic susceptibility data of a CoAs single crystal. (b) Field dependence of magnetization between $\pm$140~kOe at various temperatures. Inset: a closeup of the 2~K data between $\pm$10~kOe. The initial upsweep is red, the downsweep blue, and the return sweep light green.}
    \label{fig:Figure2}
\end{figure}

Single crystal heat capacity data were taken at low temperature. Figure 1(b) shows the data in zero field with a straight line fit to the standard low temperature heat capacity model $C/T = \gamma{}~+~\beta{}T^{2}$, for 4.5~K~$<$~\textit{T}~$<$~7~K, where the former and latter terms represent the electron and phonon contributions, respectively. The fit yields a Sommerfeld coefficient $\gamma{}$~=~6.91~$\mathrm{\frac{mJ}{K^2~mol}}$ and, from $\beta$, a Debye temperature $\theta{}_D$~=~397~K. These values are close to those seen in FeAs ($\gamma = 6.652~\mathrm{\frac{mJ}{K^2~mol}}$, $\theta_{D}~=~353~$K)\cite{Gonzalez-AlvarezFeAs} and CrAs ($7.5~\mathrm{\frac{mJ}{K^2~mol}}$ and 370~K)\cite{BlachnikCrAsHC}. For MnP, $\gamma$ is estimated to be 5.4\textendash{}7.6~$\mathrm{\frac{mJ}{K^2~mol}}$, with large uncertainty due to magnetic contributions\cite{TakaseMnPHC}. The closeness of these values shows the electronic and phononic similarities of those compounds with CoAs. Closer to 1.8~K, we see a subtle bump followed by a drop in $C/T$. Similar behavior was observed in near ferromagnets CaNi$_2$ and CaNi$_3$\cite{JescheCaNi}, and a low temperature enhancement in $C/T$ is a known indicator of spin fluctuations\cite{DoniachHCEnhancement, TrainorUAl2HCEnhancement}. In CoAs this feature is much less noticeable. However, it is still present in measurements in fields up to 14~T, where there is a slight positive deviation from linearity in \textit{C/T} below 10~K$^{2}$, followed by a lower temperature drop [Fig.~1(b), inset]. The survival of this lobe feature further supports possible spin polarization.

Measurements of magnetization were also done on CVT crystals using both the vibrating sample magnetometer option in a 14~T Quantum Design DynaCool Physical Properties Measurement System and a 7~T SQUID Magnetic Properties Measurement System. CoAs has a small moment that increases slightly as temperature is initially decreased, with a broad peak around 225~K [Fig.~2(a)], followed by a minimum near 35~K. The single crystal susceptibility $\chi{}$, including the extrema, is generally similar in appearance and magnitude to previous polycrystal reports\cite{SelteCoAs, SaparovTAs}, though the minimum at low temperatures is not as sharp as the kink seen by Saparov \textit{et al.}, and in our case the low temperature susceptibility does not exceed the higher temperature value.  A similar broad peak is also seen in FeAs\cite{SegawaFeAs} just above 200~K, which has a similar overall shape of temperature-dependent susceptibility above its 70~K spin density wave onset to CoAs. Motizuki qualitatively explained the observed maxima in both compounds as stemming from the temperature dependent spin fluctuations, which saturate in amplitude above the temperature of the peak\cite{MotizukiChiModel}. Thus, like the low temperature bump in heat capacity, the susceptibility peak in CoAs indicates the presence of significant spin fluctuations.

At low fields, the field cooled curve shows a larger low temperature upturn than the zero field cooled curve, but the difference is small and disappears above 2~kOe. The size of the upturn is also sample dependent, so we conclude that it is the result of paramagnetic impurities, without which $\chi{}$ would simply plateau with further temperature decrease. The appearance of the temperature-dependent data, including positions of the local extrema, did not vary with the magnitude of applied field between 0.5 and 70~kOe, nor did the susceptibility values. Field-dependent magnetization up to 140~kOe [Fig.~2(b)] is nonsaturating at all temperatures from 300~K to 2~K, and nonlinear at low field for T~$\leq{}$10~K, perhaps from the increasing contribution of impurities. Despite this, there is negligible hysteresis at 2~K [Fig.~2(b), inset], indicating a lack of clear ferromagnetism. An Arrott plot using M(H) data at 2~K similarly shows no sign of long range magnetic ordering. All signs point to CoAs being paramagnetic, the same conclusion reached in previous magnetic susceptibility and 4.2~K neutron diffraction measurements\cite{SelteCoAs, SaparovTAs}. A weak FM moment was observed in a powder sample by another group\cite{LewisB31}, but could be attributed to the inclusion of $^{57}$Fe in those samples for later M\"{o}ssbauer study.

\section{\label{sec:Theory}Theoretical Calculations}

\begin{figure*}
    \centering
    \includegraphics[width= 1\textwidth]{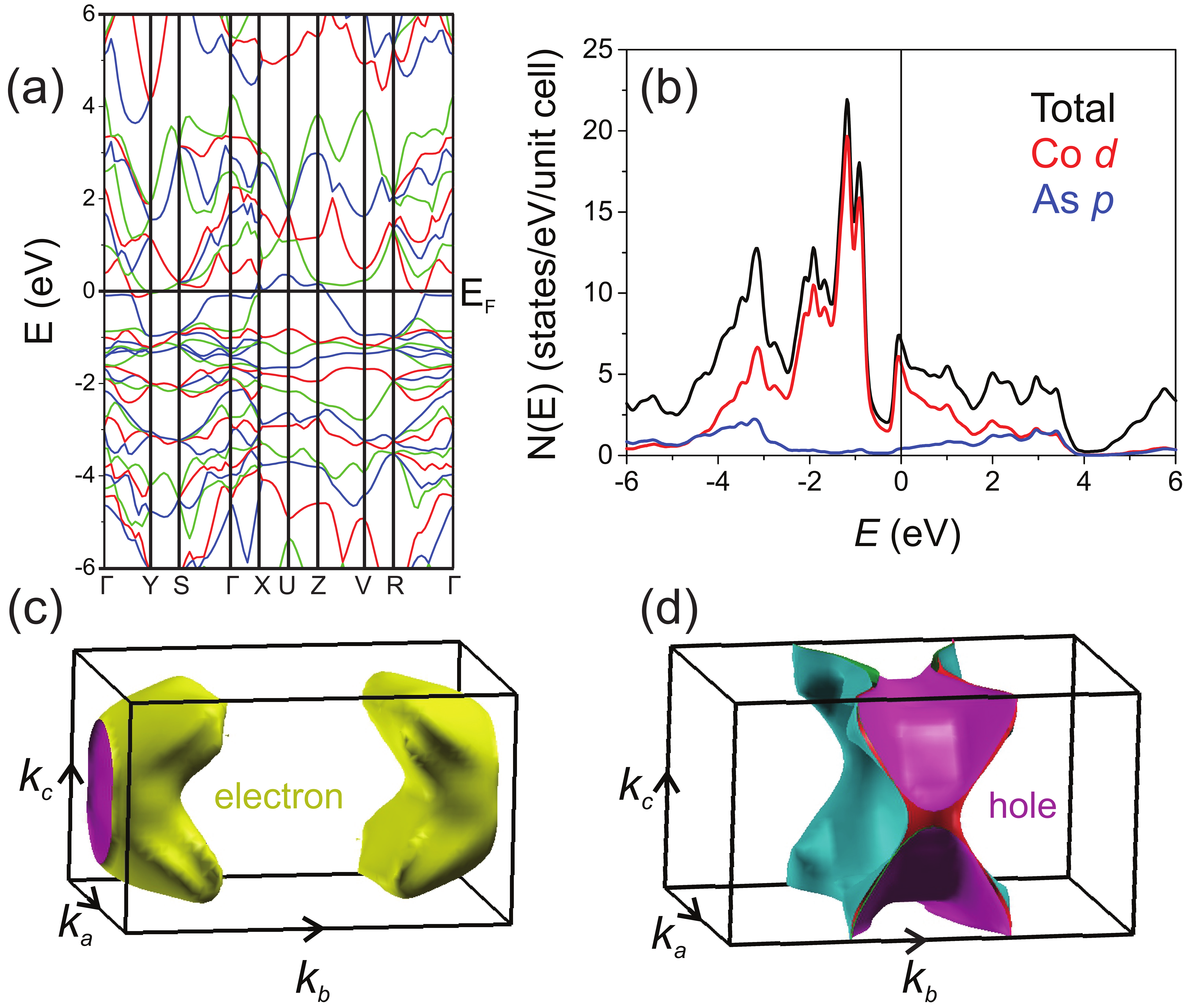}
    \caption{DFT-calculated (a) band structure (where different colors distinguish different bands), (b) density of states, (c) electron and (d) hole Fermi surfaces of paramagnetic CoAs.}
    \label{fig:Figure3}
\end{figure*}

We obtained the electronic structure of CoAs via first-principles density functional theory (DFT) calculation of the paramagnetic state. The calculation was conducted using the WIEN2K\cite{SchwarzWIEN2K} implementation of the full potential linearized augmented plane wave method within the PBE generalized gradient approximation using the lattice parameters obtained from powder XRD. The k-point mesh was taken to be 11~$\times$~17~$\times$~10. Figure~3 shows the paramagnetic band structure, density of states and Fermi surface of CoAs. The Fermi surface consists of two hole pockets and two electron pockets, and the bands around the Fermi level are dominated by the Co \textit{d} orbitals. The electron pockets have a ``Czech hedgehog'' shape centered at the \textit{Y} point. The concentric hole pockets occupy the center of the first Brillouin zone but spread into other zones, resembling connected hourglasses.

Four possible magnetic ground states were considered: paramagnetic, ferromagnetic, and two distinct antiferromagnetic orderings\textemdash{}one in which Co atoms align ferromagnetically with nearest neighbors and antiferromagnetically with next nearest neighbors, and another where they are antiferromagnetic with both. DFT results show a preference for FM over PM in CoAs by 20~meV/Co atom, with the two AFM scenarios at much higher energies. This energy difference translates to about 230~K, the location of the local maximum in susceptibility. The calculated moment for the FM state is 0.28$\mu_B$, where $\mu{}_B$ is a Bohr magneton. As previously noted, measurements here and in other works show no indication of long range magnetic ordering in CoAs. Additionally, the magnetization at 140~kOe and 2~K is still two orders of magnitude smaller than the expected moment. Neutron diffraction measurements saw no purely magnetic reflections and set an upper limit of 0.1$\mu{}_B$ at 4.2~K on any potential FM moment\cite{SelteCoAs}. The energy difference between ferro- and paramagnetism is small. Combined with experimental results, it is possible that PM is in fact the lowest energy state, or that the onset of magnetic ordering occurs at an even lower temperature than that reached in these experiments or previous ones.

Motivated both by calculations and physical property measurements, we considered possible near ferromagnetism in CoAs. One way to quantify this is the dimensionless Wilson ratio R$_W$~=~$\frac{4\pi{}^2k_B^2\chi_0}{3\mu{}_0(g_e\mu{}_B)^2\gamma{}}$, where k$_B$ is the Boltzmann constant, $\chi_0$ the 0~K spin susceptibility, $\mu_0$ the permeability of free space, and $g_e$ the electron g-factor. R$_W$ is unity for a free electron gas, and much larger values indicate proximity to a FM instability. For CoAs, R$_W$~=~6.2, comparable to the values for the known near ferromagnet Pd (R$_W$ = 6\textendash{}8)\cite{QuWilsonRatio} and BaCo$_2$As$_2$ (7\textendash{}10, depending on field orientation)\cite{SefatBaCo2As2}, which is thought to be near a magnetic quantum critical point. R$_W$ can be reexpressed as the Stoner factor Z~=~1-$\frac{1}{\text{R}_W}$, where Z~$\rightarrow$~1 signifies stronger ferromagnetic correlations. Z$_{CoAs}$~=~0.84, similar to near ferromagnets CaNi$_2$ (0.79) and CaNi$_3$ (0.85)\cite{JescheCaNi}, which showed a low temperature enhancement in C/T. Based on both theoretical and experimental results, we suspect that there exist low temperature FM fluctuations in CoAs. Future work with chemical substitution or applied pressure may stabilize a magnetic state.

We have also made DFT calculations for paramagnetic FeAs with the same methods and present a comparison to CoAs in Fig.~4. The electronic structure between PM FeAs and CoAs differs only by a rigid band shift, as demonstrated by the fact that raising the Fermi level of the calculated FeAs band structure [Fig.~4(a)] and density of states plot [Fig.~4(b)] nearly reproduces the CoAs equivalent in both cases. The shift is about 1~eV, which is logical given that Co has an extra electron compared to Fe. To explore this relationship further we recalculated the PM FeAs band structure using the CoAs lattice parameters. It should be noted that while \textit{a} and \textit{c} are smaller for CoAs compared to FeAs, \textit{b} is actually longer. There is a negligible difference in PM FeAs band structure calculated using the two unit cell sizes, indicating that the 70~K spin density wave onset in FeAs and corresponding lack of ordering in CoAs has a more complicated origin than just unit cell size and bond distance.

The predicted PM Fermi surface of CoAs is very different from that of FeAs\cite{ParkerFeAs}, and low temperature AFM is also highly unfavored in CoAs. The relatively empty dispersion in the region between -1 and 0~eV and significant dropoff in the density of states at E$_{F}$ are probably responsible for this, as the density of states near E$_F$ is thought to have a large impact on the behavior of spin fluctuations\cite{MotizukiChiModel}. A significant difference in magnetic ordering has been seen in other Fe and Co binaries. FeSe and CoSe can both be synthesized in a tetragonal structure. FeSe is a PM, potentially spin-fluctuation mediated superconductor (T$_c$~=~8~K)\cite{HsuFeSe}, while below 10~K CoSe is a spin glass\cite{WilfongCoSe}. Like the arsenides, their band structures and densities of states have essentially a 1~eV shift between them\cite{ZhouCoCh}. CoSb (in the hexagonal NiAs structure) goes from PM to a spin glass phase with Fe substitution, while FeSb itself is AFM\cite{AmornpitoksukCoSb}. For both pnictides and selenides of Fe and Co, rigid band shifts have a large effect on ground state magnetism.

\begin{figure}
    \centering
    \includegraphics[width=0.48\textwidth]{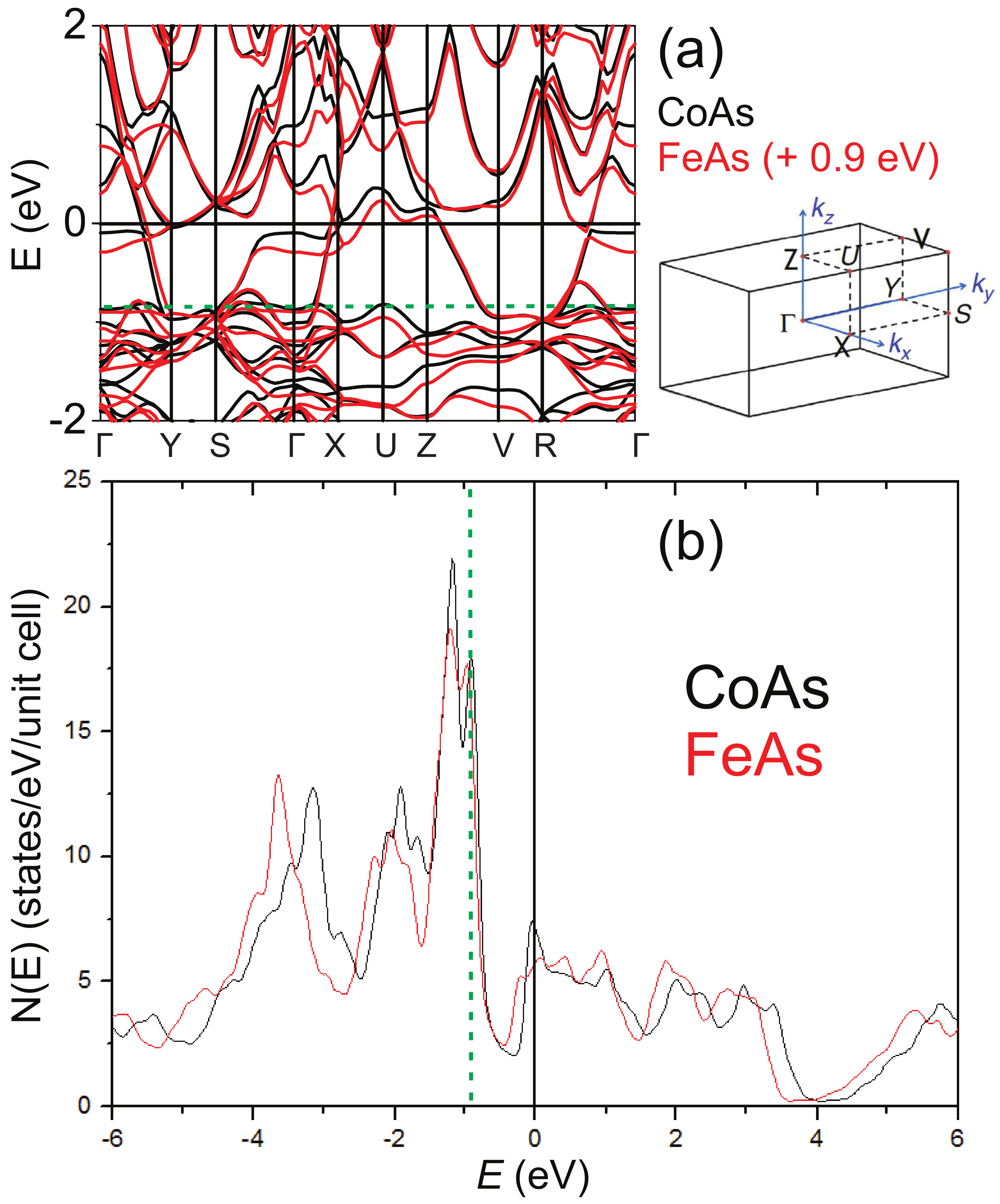}
    \caption{(a) A comparison of the band structures of paramagnetic FeAs (after having its Fermi level shifted up by 0.9~eV) and CoAs, with a schematic of the Fermi surface convention used. (b) A density of states comparison, with FeAs again shifted by 0.9~eV. In both plots a dashed green line indicates the original Fermi level of FeAs.}
    \label{fig:Figure4}
\end{figure}

\section{\label{sec:QOs}Quantum Oscillations}

Quantum oscillations emerge due to the creation of Landau levels (LL), quantized energy levels in the density of states that form when a sample is placed in a magnetic field. As field strength changes, the energy of a LL does as well. If a LL passes through the chemical potential there is a change in its occupation, producing an oscillatory effect. These oscillations are observable in various physical properties, but the two most commonly measured are resistance and magnetization, in which case they are known as Shubnikov-de Haas and de Haas-van Alphen oscillations, respectively. The oscillation frequency is directly proportional to the cross sectional area of the pocket around which a carrier makes a cyclotron orbit perpendicular to the applied field. Analysis of oscillation frequency and amplitude as a function of angle, temperature, and field strength gives information about the Fermi surface\cite{ShoenbergOscillations}.

Measurements of longitudinal resistance and magnetic torque were made at the DC Field Facility of the National High Magnetic Field Laboratory in Tallahassee, Florida using the 31.5~T, 50~mm bore magnet on single crystals grown from Bi flux. A typical four wire setup was used for magnetotransport and piezoresistive cantilevers for torque, with both attached to a rotating probe in a He-3 system with a base temperature of 400~mK. Magnetotransport was featureless and small, showing $H^2$ dependence without oscillations. However, oscillations were readily observable in the more sensitive torque signal as low as 6~T. Figure~5(a) shows the torque at selected orientations of the sample relative to applied field. Various oscillation frequencies emerge in the data over the entire angular range, though some correspond to harmonics or the sum of independent fundamental frequencies. To compare experimental results to band structure predictions, we generated theoretical quantum oscillation frequencies and effective masses from our DFT calculations using the Supercell K-space Extremal Area Finder (SKEAF) program\cite{RourkeSKEAF}. 

\begin{figure*}
    \centering
    \includegraphics[width=1\textwidth]{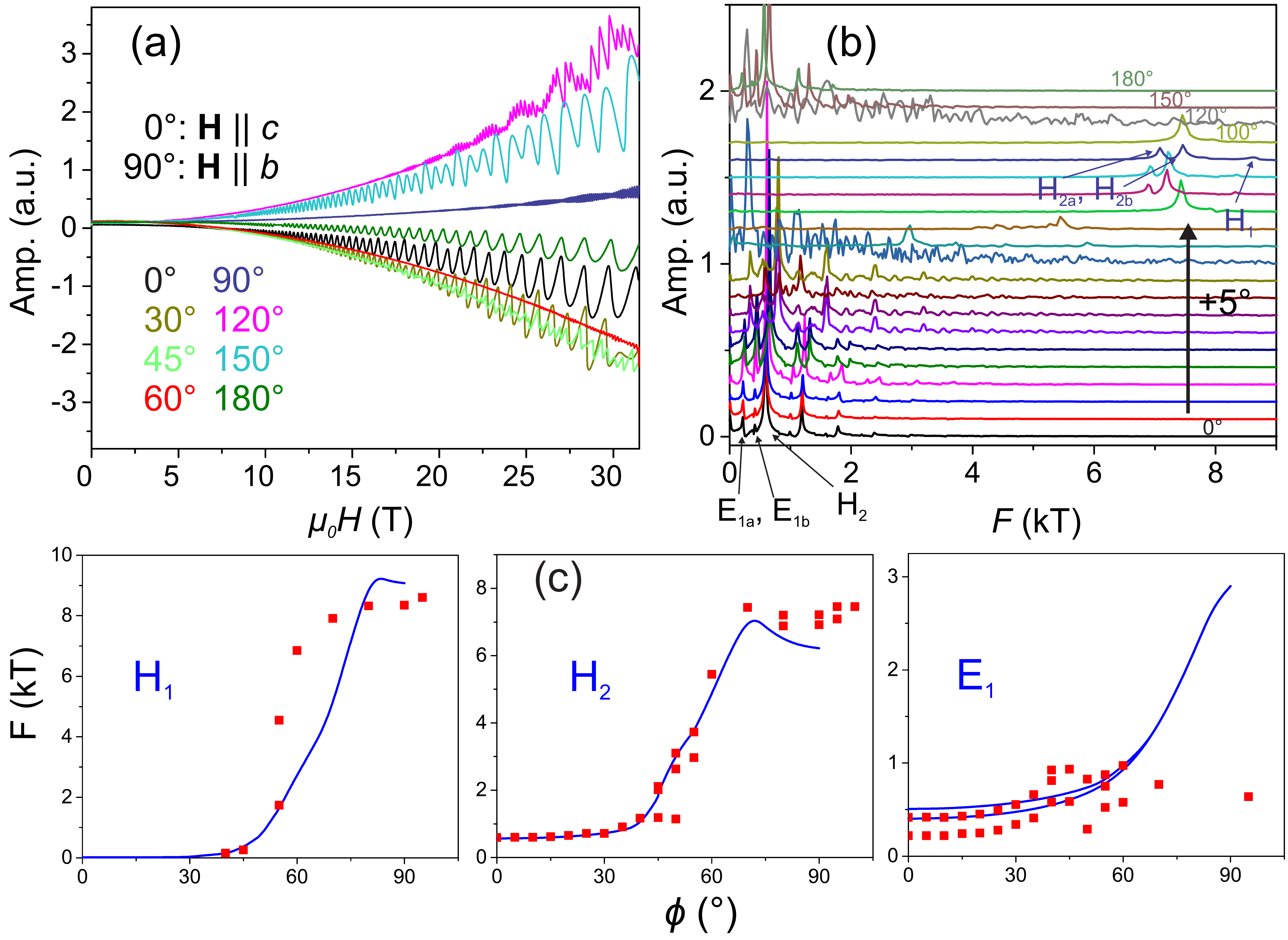}
    \caption{(a) Magnetic torque of a CoAs single crystal at various angles at base temperature (400-500~mK), where \textbf{H}~$\parallel$~\textit{c} at 0\degree{} and $\parallel$~\textit{b} at 90\degree{}. (b) Fast Fourier transforms of the oscillatory component of the torque for each angle, offset for clarity. Unlabeled peaks are either higher harmonics or sums of the fundamental frequencies. Data between 0\degree{} and 100\degree{} were taken in 5\degree{} increments. (c) A comparison of observed oscillation frequencies (red squares) and those expected from the paramagnetic DFT band structure (blue lines) for three of the four predicted oscillation bands. H$_1$ and H$_2$ are hole bands, E$_1$ is an electron band.}
    \label{fig:Figure5}
\end{figure*}

\subsection{\label{sec:Angle}Fermi Surface Geometry}

The change in oscillation frequency spectrum with applied field angle reflects the geometry of the Fermi surface pockets. In our measurement $\phi$~=~0\degree{} corresponded to \textbf{H}~$\parallel$~\textit{c}, while $\phi$~=~90\degree{} was \textbf{H}~$\parallel$~\textit{b}. The \textit{c} axis was confirmed by single crystal XRD on the platelike sample. Given the ambiguity of the orthorhombic structure for a square plate crystal, the \textit{b} axis determination was made on the basis of comparison to predictions generated by the SKEAF program. The observed angular dependence matched very well to predictions for \textbf{H}~$\parallel$~\textit{b} but not at all for \textbf{H}~$\parallel$~\textit{a}. Data were taken in 5\degree{} intervals from 0\degree{} to 100\degree{} and at 120\degree{}, 150\degree{}, and 180\degree{} [Fig. 5(b)].

There were four distinct frequencies predicted by SKEAF, but in the experimental data we only observed three independent sets of oscillations. The three can be assigned to one of the predicted electron bands and the two hole bands based on similarities between calculated and observed frequencies [Fig.~5(c)]. There is some slight disagreement in exact frequency value, but the angular dependence matches well. The unobserved band corresponds to a second electron band, and does not match any of the angular dependent data. It is not atypical for a predicted band to be absent in measurements\cite{RourkeSKEAF}. In our case, the missing band has the highest predicted effective mass, which would reduce the oscillation amplitude and make it more difficult to detect. Additionally, the predicted frequencies correspond to cross sectional areas much larger than the first Brillouin zone, and an erratic angular dependence indicative of a potentially unrealistic orbit. More important is that all experimentally observed frequencies can be indexed to a theoretical band. The extremal orbits predicted around the electron and hole pockets all have a nontrivial shape, and correspondingly the expected oscillation frequency also varies widely with angle.

The fact that, in spite of this complexity, we do see agreement in the plane in which rotational measurements were done indicates that the paramagnetic Fermi surface in Fig.~3 reflects the true CoAs Fermi surface as far as we are able to determine. We refer to the three bands as H$_1$, H$_2$, and E$_1$ with H and E denoting hole and electron, respectively. The observed hole pockets show consistently higher oscillation frequencies than the electron pocket, meaning that they are larger and explaining the positive Hall coefficient. The linearity of the Hall signal, despite the presence of multiple carriers, is presumably due to a much greater number of hole carriers indicated by the larger oscillation frequencies of the hole bands. Carrier concentration \textit{n} scales as with oscillation frequency as  $F^{1.5}$, so frequencies roughly three times as large for the two hole bands compared to the one electron band would mean $n_h$ is an order of magnitude larger than $n_e$.

To extract oscillation frequencies a third order polynomial was fit to the raw data and subtracted to obtain a residual oscillatory signal. Figure~6(a) gives examples at 100\degree{} and 180\degree{}. A fast Fourier transform (FFT) was then performed on the residue, with the final products shown in Fig.~5(b). The Onsager relation equates the oscillation frequency and area enclosed by the cyclotron orbit\cite{ShoenbergOscillations}: $F~=~\frac{\hbar{}}{2\pi{}e}A$. The oscillation frequencies, and therefore Fermi surface cross sectional areas, increase as the field more closely aligns with the \textit{b} axis. Accordingly, both the electron and hole Fermi surfaces have their largest cross sections in the \textit{ac} plane [Fig.~3]. Figure~5(c) shows the frequency at which peaks were observed at different angles (red squares), as well as the angular dependence predicted via SKEAF (blue lines) in the range 0\degree{}\textendash{}90\degree{}. E$_1$ shows multiple peaks in this range. This and the increase in frequency closer to 90\degree{} are both in line with theory. The increase in frequency also comes with a decrease in amplitude, and it is not until 70\degree{} that peaks are again clear. At this point only H$_1$ and H$_2$ are observed, in the 7\textendash{}8~kT range, with intermittent lower peaks potentially corresponding to E$_1$.

There are two angular ranges for which we do not observe three bands: H$_1$ does not appear for 0\degree{}~$<\phi<$~40\degree{}, and E$_1$ does not appear at almost all angles above 60\degree{}. In both cases the disappearance of frequencies can be explained by experiment-related factors, rather than disagreement with theory. Until 40\degree{} predicted H$_1$ frequencies are less than 35~T, and such a low frequency makes them hard to pick out in our FFTs, which cover a large, higher frequency range. At high angles, the predicted effective mass for E$_1$ increases, exceeding 4$m_e$. This will decrease oscillation amplitude, and so it is unsurprising to see this frequency band become less prominent in the data.

\subsection{\label{sec:LK}Effective Mass}

The decrease in oscillation amplitude with temperature can be used for the average effective mass $m^*$ in a given orbital plane via the Lifshitz-Kosevich factor $R_T~=~\frac{\alpha{}m\text{*}T/(\mu{}_0Hm_e)}{\text{sinh}(\alpha{}m\text{*}T/(\mu{}_0Hm_e))}$ where $\alpha = 2\pi^2ck_B/e\hbar \approx 14.69$ T/K, \textit{c} is the speed of light, $e$ the electron charge, and $\hbar$ the reduced Planck constant\cite{ShoenbergOscillations}. Temperature dependence for CoAs was taken at two different angles: 180\degree{} and 100\degree{}. That is, at \textbf{H} $\parallel$ \textit{c} and near \textbf{H} $\parallel$ \textit{b}, respectively. Data were not taken exactly at 90\degree{} due to reduced torque amplitude along the crystal axis. At these angles, by fitting amplitude as a function of temperature to the LK formula we obtain an average carrier mass for extremal orbits in the $k_a$\textendash{}$k_c$ and $k_a$\textendash{}$k_b$ planes.

At 100\degree{} [Fig.~6(a), left] three frequencies appear: H$_1$ and two split peaks stemming from the H$_2$ band. While the amplitudes show a decay with temperature, the fits to the LK formula are not great. Nevertheless we obtain $m_{H1,~exp.}^*$~=~2.3~$m_e$, $m_{H2a,~exp.}^*$~=~2.6$m_e$, and $m_{H2b,~exp.}^*$ = 2.4$m_e$. The subscripts \textit{a} and \textit{b} denote the lower and higher of the two split frequencies. Theoretical predictions gave $m_{H1,~th.}^*$~=~2.99$m_e$ and $m_{H2,~th.}^*$~=~1.91$m_e$ at 90\degree{}\textemdash{}the frequency splitting we see in H$_2$ was not predicted. At 180\degree{} [Fig.~6(a), right] two split E$_1$ frequencies as well as one H$_2$ frequency are seen. Here the effective mass fits are much better [Fig.~6(b), right], and the H$_2$ signal survives to at least 15 K, indicative of lighter masses at this angle: $m_{E1a,~exp.}^*$~=~0.70$m_e$, $m_{E1b,~exp.}^*$~=~0.36$m_e$, and $m_{H2,~exp.}^*$~=~0.46$m_e$, compared to predicted values of $m_{E1a,~th.}^*$~=~1.43$m_e$, $m_{E1b,~th.}^*$~=~1.44$m_e$, and $m_{H2,~th.}^*$~=~0.33$m_e$. Overall the predicted and observed masses do not show close agreement. Encouragingly, however, the prediction of smaller masses for \textbf{H}~$\parallel$~\textit{c} compared to \textbf{H}~$\parallel$~\textit{b} is borne out by the data. The effective masses are all generally close to the free electron mass, indicating a lack of significant correlated behavior.

The effective mass gives us an alternate method of calculating the Sommerfeld coefficient, as $\gamma$~=~$\frac{\pi{}k_B^2N_A}{3E_F}$. $N_A$ is the Avogadro number and $E_F$~=~$\frac{\hbar{}^2k_F^2}{2m^*}$ the Fermi energy, with $k_F^2$ equal to the cross sectional Fermi surface area perpendicular to the direction of magnetic field. At a specific angle, a contribution proportional to $m^*$ from each observed orbit is added together to get the total $\gamma$. If the quantum oscillations result is significantly smaller than the value obtained from heat capacity measurements, it means that the data are missing a number of carriers at the Fermi level, potentially from an unobserved additional Fermi surface pocket. For \textbf{H} $\parallel$ \textbf{[010]}, the total $\gamma$ comes out to 3.9~$\mathrm{\frac{mJ}{K^2~mol}}$, close to value of 6.83~$\mathrm{\frac{mJ}{K^2~mol}}$ we directly measured in heat capacity experiments. For \textbf{H} $\parallel$ \textbf{[001]}, $\gamma$~=~16.9~$\mathrm{\frac{mJ}{K^2~mol}}$, a much larger value due to the reduced effective masses in the \textit{ab} plane. In either case we do not appear to be missing any contributions at the Fermi level, despite not observing the fourth DFT-predicted band.

\begin{figure}
    \centering
    \includegraphics[width=0.47\textwidth]{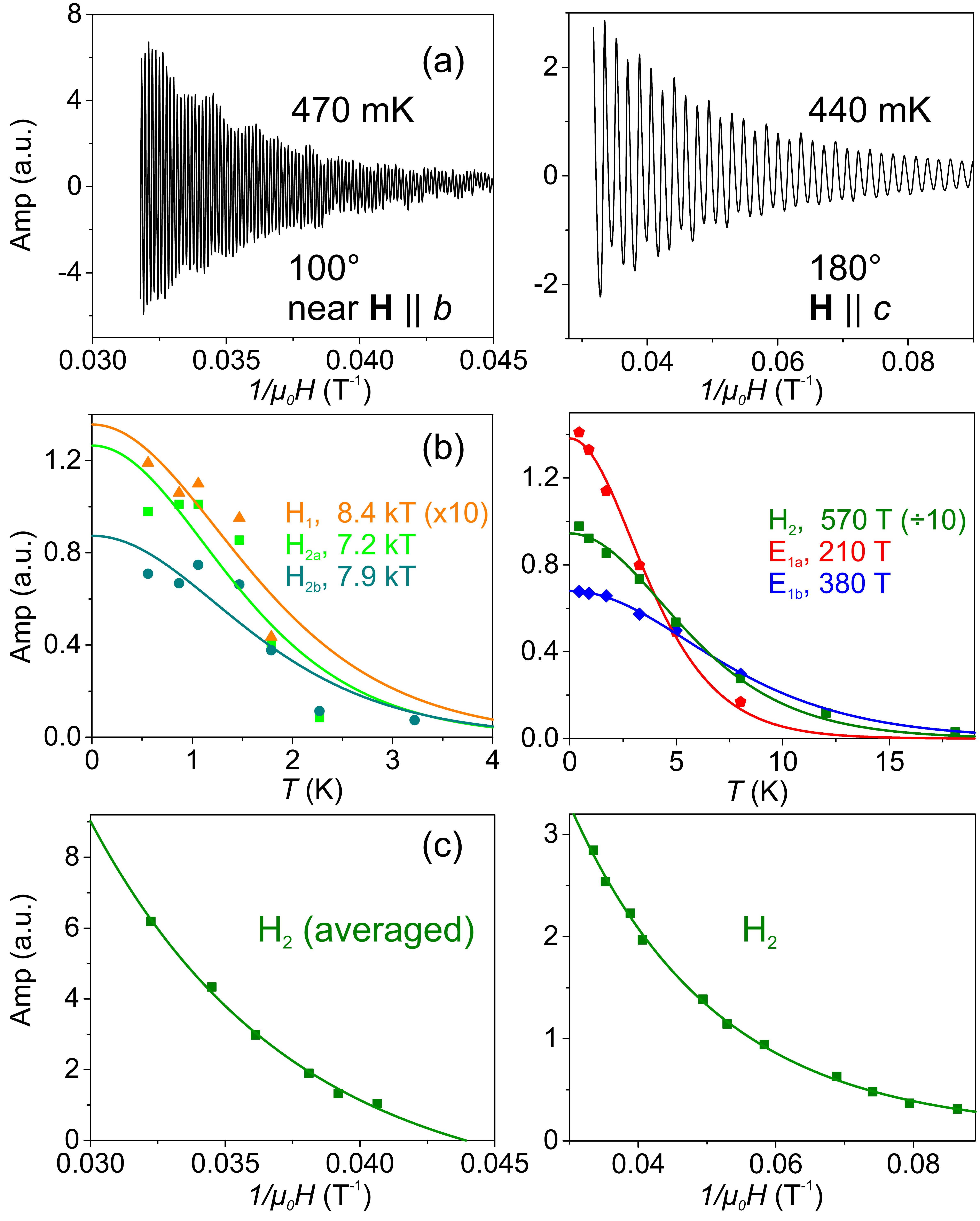}
    \caption{(a) Residual oscillatory signal for 100\degree{} (near \textbf{H}~$\parallel$~\textit{b}) and 180\degree{} (\textbf{H}~$\parallel$~\textit{c}). (b) Temperature dependence of the amplitude of the observed peaks at both angles, with fits to the Lifshitz-Kosevich factor. To ease comparison between different bands, on the left H$_1$ amplitude is increased by a factor of 10 and on the right H$_2$ decreased a factor of 10. (c) A plot of the peak amplitude versus inverse field for H$_2$ at both angles, with accompanying exponential decay fits to solve for the Dingle temperature. Only H$_2$ was used since it was the most prominent frequency at both angle. Data for 100\degree{} are an average of the two observed peaks.}
    \label{fig:Figure6}
\end{figure}

\subsection{\label{sec:Dingle}Dingle Temperature}

Oscillation amplitude should decrease in an exponential ``envelope'' with inverse field, and if the effective mass is known one can solve for the Dingle temperature $T_D$ as $R_D = \text{exp}(\frac{-\alpha{}m^{*}T_D}{\mu{}_0Hm_e})$. $T_D$ can be translated to the carrier scattering rate $\Gamma$: $T_D = \frac{\hbar}{2\pi{}k_B}\Gamma{}$. One difficulty in determining the Dingle temperature is that the amplitude must be extracted from the oscillatory signal, rather than the Fourier transform. This means it is only feasible to do Dingle analysis for peaks with a large relative amplitude at an angle for which $m^*$ is known. Thus we only have $T_D$ for H$_2$ at 100\degree{} (by averaging the split peaks) and 180\degree{}, since the exponential decay of H$_2$ dominates the oscillatory signal [Fig.~6(a)]. We can then fit the position of peaks with inverse field, giving $T_{D,~H2}$~=~3.66~K and 14.1~K at 100\degree{} and 180\degree{}, corresponding to $\Gamma_{H2}$~=~3.0~$\times$~10$^{12}$~s$^{-1}$  and 1.2~$\times$~10$^{13}$~s$^{-1}$, respectively. For a spherical Fermi surface, the mean free path $\ell{} = \frac{\hbar{}k_F}{m^{*}\Gamma}$, and $k_F$ can be calculated from the cross sectional pocket area as $A = \pi{}k_F^2$. The angular dependence of H$_2$ clearly shows it is not spherical, but we can average the values for the two different field directions to obtain a rough estimate of $\ell{}~=~500~$\AA{} for H$_2$. Overall we see a large anisotropy in the hole pocket in terms of both effective mass and Dingle temperature between the \textit{ac} and \textit{ab} planes, as the values listed in Table I indicate. In the AFM states of CrAs, MnP, FeP, and FeAs, the \textit{ab} plane features two noncollinear rotating magnetic moments\cite{ChengCrAsMnP, RodriguezFeAsSDW, KallelHelimagnetism}. This could be a sign that any magnetic fluctuations in CoAs are occurring in the \textit{ab} plane.

\begin{table}
	\centering
    \caption{Parameters extracted from CoAs torque oscillations data with field applied in different directions. Note that what is called [010] actually corresponds to an angle 10\degree{} off of the \textit{b} axis. The average $m^*$ of H$_{2a}$ and H$_{2b}$ was used to calculate $T_D$ for H$_{2, ave}$.\\}
    \label{tab:Table1}

\renewcommand{\arraystretch}{1.2}
\begin{tabular}{ | c | c | c | c | c | c | c | c |}
	\hline
	Band & \textbf{H} $\parallel$ \textbf{[hkl]} & F (kT) & \textit{m*}/$m_e$ & $T_D$ (K) \\
	\hline
	H$_1$ & \textbf{[010]} & 8.39 & 2.3 & \textemdash{} \\
	\hline

	H$_{2a}$ & \textbf{[010]} & 7.20 & 2.6 & \textemdash{} \\
	\hline
	
	H$_{2b}$ & \textbf{[010]} & 7.90 & 2.4 & \textemdash{} \\
	\hline
	
	H$_{2, ave}$ & \textbf{[010]} & \textemdash{} & \textemdash{} & 3.66 \\
	\hline

	H$_{2}$ & \textbf{[001]} & 0.57 & 0.46 & 14.1 \\
	\hline

	E$_{1a}$ & \textbf{[001]} & 0.21 & 0.70 & \textemdash{} \\
	\hline
	
	E$_{1b}$ & \textbf{[001]} & 0.38 & 0.36 & \textemdash{} \\
	\hline
	
\end{tabular}

\end{table}

\section{\label{sec:Conclusion}Conclusion}

We have reported two different growth methods of single crystal CoAs, using either Bi flux or I$_2$ vapor transport. The flux-grown crystals have a lower residual resistivity and more consistent orientation, but the vapor transport samples can grow much larger, enabling bulk measurements such as magnetization or heat capacity. Data show hole-dominated metallic conduction with no indication of long range magnetic ordering down to 1.8~K in single crystal CoAs, despite predictions slightly favoring weak moment ferromagnetism. We have observed de Haas-van Alphen oscillations in torque starting from 6~T up to 31.5~T, and their angular dependence is in line with the geometry of the calculated paramagnetic Fermi surface.

Many 3\textit{d} transition metal pnictide compounds such as MnP, CrAs, FeAs, and the iron-based superconductors have shown unique magnetic arrangements and often superconductivity upon suppression of ordered magnetism. CoAs is paramagnetic, but a nonmonotonic temperature dependence of magnetic susceptibility and low temperature enhancement of heat capacity point to possible magnetic fluctuations. The notion of a ferromagnetic instability is also supported by DFT calculations, which actually favor long range ordering at zero temperature. Furthermore, there are many similarities to antiferromagnetic FeAs, both in magnetic susceptibility and electronic structure, which differs only by a one electron rigid band shift, with the effect of unit cell size apparently negligible. An earlier pressure study\cite{LymanCoAsPressure} up to 10~GPa showed a possible structural transition at 7.8~GPa, and other elements of the structure identified by the authors indicate a potential sensitivity to lattice shifts at high pressures. Future work with chemically substituting or applying pressure to binary CoAs may lead to the discovery of structural, magnetic, or superconducting transitions, similar to those exhibited by neighboring isostructural transition metal-pnictide binaries.

\section{\label{sec:Acknowledge}Acknowledgments}

The authors acknowledge useful discussion with A. Nevidomskyy. This research was supported by Air Force Office of Scientific Research award No.~FA9550-14-1-0332 and National Science Foundation Division of Materials Research award No.~DMR-1610349. A portion of this work was performed at the National High Magnetic Field Laboratory, which is supported by National Science Foundation Cooperative Agreement No.~DMR-1157490 and the State of Florida. We acknowledge the support of the Maryland NanoCenter and its FabLab.

\bibliography{CoAsPaper}

\end{document}